\documentclass[letterpaper,twocolumn,prl,aps,superscriptaddress,floatfix]{revtex4}

\usepackage[latin9]{inputenc}
\usepackage{amsmath}
\usepackage{amssymb}
\usepackage{graphicx}
\usepackage{esint}
\usepackage{times}
\usepackage[markup=underlined]{changes}

\usepackage[unicode=true,
bookmarks=true,bookmarksnumbered=false,bookmarksopen=false,
breaklinks=false,pdfborder={0 0 1},backref=false,colorlinks=true]
{hyperref}
\hypersetup{
    linkcolor=magenta, urlcolor=blue, citecolor=blue, pdfstartview={FitH}, hyperfootnotes=false, unicode=true}

\makeatletter

\pdfpageheight\paperheight
\pdfpagewidth\paperwidth

\@ifundefined{textcolor}{}
{%
    \definecolor{BLACK}{gray}{0}
    \definecolor{WHITE}{gray}{1}
    \definecolor{RED}{rgb}{1,0,0}
    \definecolor{GREEN}{rgb}{0,1,0}
    \definecolor{BLUE}{rgb}{0,0,1}
    \definecolor{CYAN}{cmyk}{1,0,0,0}
    \definecolor{MAGENTA}{cmyk}{0,1,0,0}
    \definecolor{YELLOW}{cmyk}{0,0,1,0}
}

\usepackage{xcolor}
\usepackage{soul}

\newcommand{\ket}[1]{\ensuremath{\left|#1\right\rangle}}
\definecolor{blue}{rgb}{0,0,1}
\definecolor{red}{rgb}{0,0,0}
\definecolor{green}{rgb}{0,1,0}
\newcommand{\red}[1]{\textcolor{red}{ #1}}

\usepackage{soul}

\makeatother

\begin{document}
%\title{Scalable quantum metrology using a confocal system in Fock space}
%\title{Deterministic quantum metrology with photon-number squeezing of 500 photons}
\title{Quantum Confocal Microscopy in Fock Space with a 19~dB Metrological Gain}

\author{Ziyue Hua}
\thanks{These authors contributed equally to this work.}
\author{Chuanlong Ma}
\thanks{These authors contributed equally to this work.}
\author{Yilong Zhou}
\thanks{These authors contributed equally to this work.}
\author{Yifang Xu}
\affiliation{Center for Quantum Information, Institute for Interdisciplinary Information Sciences, Tsinghua University, Beijing 100084, China}

\author{Zi-Jie Chen}
\author{Weizhou Cai}
\affiliation{Laboratory of Quantum Information, University of Science and Technology of China, Hefei 230026, China}

\author{Jiajun Chen}
\author{Lintao Xiao}
\author{Hongwei Huang}
\author{Weiting Wang}
\affiliation{Center for Quantum Information, Institute for Interdisciplinary Information Sciences, Tsinghua University, Beijing 100084, China}

\author{Hekang Li}
\affiliation{School of Physics and ZJU-Hangzhou Global Scientific and Technological Innovation Center, Zhejiang University, Hangzhou 310027, China}
\affiliation{Hefei National Laboratory, Hefei 230088, China}
\author{Haohua Wang}
\affiliation{School of Physics and ZJU-Hangzhou Global Scientific and Technological Innovation Center, Zhejiang University, Hangzhou 310027, China}
\affiliation{Hefei National Laboratory, Hefei 230088, China}

\author{Ming Li}
\email{lmwin@ustc.edu.cn}
\affiliation{Laboratory of Quantum Information, University of Science and Technology of China, Hefei 230026, China}
\affiliation{Hefei National Laboratory, Hefei 230088, China}

\author{Chang-Ling Zou}
\email{clzou321@ustc.edu.cn}
\affiliation{Laboratory of Quantum Information, University of Science and Technology of China, Hefei 230026, China}
\affiliation{Hefei National Laboratory, Hefei 230088, China}

\author{Luyan Sun}
\email{luyansun@tsinghua.edu.cn}
\affiliation{Center for Quantum Information, Institute for Interdisciplinary Information Sciences, Tsinghua University, Beijing 100084, China}
\affiliation{Hefei National Laboratory, Hefei 230088, China}

%\date{\today}

\begin{abstract}
\textbf{Quantum metrology promises measurement precision beyond classical limits by exploiting large-scale quantum states, yet realizing this advantage faces two fundamental challenges: the deterministic preparation of non-trivial quantum probes and the efficient extraction of metrological information in high-dimensional Hilbert spaces. Here, we introduce quantum confocal microscopy in Fock space that simultaneously resolves both challenges. Drawing a direct analogy between classical wave optics and quantum state evolution in a bosonic mode, we construct a confocal system with two Fock-space lenses. The first lens deterministically focuses a coherent state into a quantum probe with a tightly concentrated photon-number distribution, while the second lens maps the metrological information back to the vacuum state for efficient readout. Using a superconducting circuit QED platform, we prepare focused probe states with mean photon numbers up to ${N} = 500$, achieving a 21.5$\pm$1.1~dB compression of the photon-number uncertainty relative to a coherent state, with a scalable quantum circuit of $\mathcal{O}(1)$ operational depth. We demonstrate a displacement sensitivity scaling as $N^{-0.416}$, approaching the Heisenberg scaling ($N^{-0.5}$), and achieve a record metrological gain of 19.06$\pm$0.13~dB beyond the standard quantum limit. This work establishes quantum confocal microscopy as a scalable and practical framework for quantum-enhanced precision measurement, readily extendable to other bosonic platforms and high-dimensional quantum many-body systems.}
\end{abstract}
\maketitle
\vskip 0.5cm

\section{Introduction}
%First paragraph: \textcolor{red}{Introduction on quantum metrology, bosonic mode}
\noindent The enhancement of measurement precision holds profound significance across diverse scientific and technological domains, ranging from the discovery of new physics to the foundation of international standards~\cite{Schnabel2010GravitationalwavesQM,Backes2021darkmatter,Wang2026darkmatter}. Quantum metrology, which leverages the superposition and entanglement of large numbers of microscopic particles or excitations, is expected to significantly improve measurement precision beyond the standard quantum limits (SQLs)~\cite{Giovannetti2004SQL,Giovannetti2011quantummetrology,Jiahao2024review,Lin2025review,Victor2025review,Chiranjib2025review}. Two complementary platforms have been exploited to validate the principles of quantum metrology. In ensembles of two-level systems, collectively entangled quantum states, such as spin-squeezed states~\cite{Cox2016DeterministicSpinSqueeze,Hosten2016SQL18dB}, GHZ states~\cite{Leibfried2004GHZmetrology,Pezze2018review}, and spin-motion entanglement states~\cite{Greve2022Entanglement,Tian-Wei2023echoingspinsqueezing}, have demonstrated quantum-enhanced measurement precision. Bosonic modes offer a distinct advantage of demonstrating large-scale quantum properties with reduced hardware complexity by exploiting the infinite-dimensional Hilbert space of harmonic oscillators~\cite{Zurek2001SubPlanck,Pirandola2018photonicquantumsensing}. The metrological capability of bosonic modes has been investigated utilizing Fock  states and their superpositions~\cite{McCormick2019ion0N,Wang2019Heisenberglimited,Wolf2019MotionalFock,Wang2022radiometryQEC,Deng2024metrology100Fock}, cat states~\cite{Vlastakis2013Cat100Photon,Xiaozhou2025VersatileMetrology,Pan2025Quantum-enhanced}, and NOON states~\cite{Nagata2007NOON,Slussarenko2017NOON}, positioning the bosonic modes as a promising platform for practical quantum-enhanced sensing.

%Second paragraph: \textcolor{red}{The Unsolved Bottleneck and The Gap}
Despite these advances, two fundamental challenges obstruct the practical realization of large metrological gains in macroscopic quantum systems. The first is state preparation. While metrological precision improves with the number of particles or excitations $N$, the preparation of high non-trivial quantum probe states becomes exponentially more demanding as $N$ grows~\cite{Escher2011noisyquantummetrology,Demkowicz2012elusiveHeisenberglimit}. In bosonic modes, for instance, photon injection~\cite{Hofheinz2008Fock,Wang2008Fock,Hofheinz2009Focksuperposition,Chu2018acousticFock} and pulse engineering~\cite{Khaneja2005GEAPE,Heeres2015SNAP,Krastanov2015SNAPUniversal,Heeres2017Grape,Eickbusch2022ECD}, require increasingly complex quantum circuits whose depth and gate count scale with $N$, making them highly susceptible to decoherence and practical imperfections. Non-deterministic approaches relying on ancilla qubit measurements can reach 100 photons~\cite{Wang2017CoherentPS,Deng2024metrology100Fock}, but their success probability {dramatically} drops with increasing $N$. 
The second bottleneck is information extraction. Even when such large-scale quantum states can be prepared, the efficient extraction of metrological information from these complex, high-dimensional probe states after interrogation remains challenging. One possible approach is to construct a time-reversal evolution of the system, either through Hamiltonian inversion ~\cite{Davis2016Approaching,Colombo2022Time-reversal,Zeyang2023quantumscramblingmetrology} or echo-type processes~\cite{Hosten2016SQL18dB,Gilmore2021ionmetrology,Tian-Wei2023echoingspinsqueezing}, which maps the probe state back near a simple reference state for efficient readout. However, such approaches demand precise control over the Hamiltonian of the entire quantum system, placing stringent and often impractical requirements on experiment. 

\begin{figure*}
    \centering
    \includegraphics[width=0.9\textwidth]{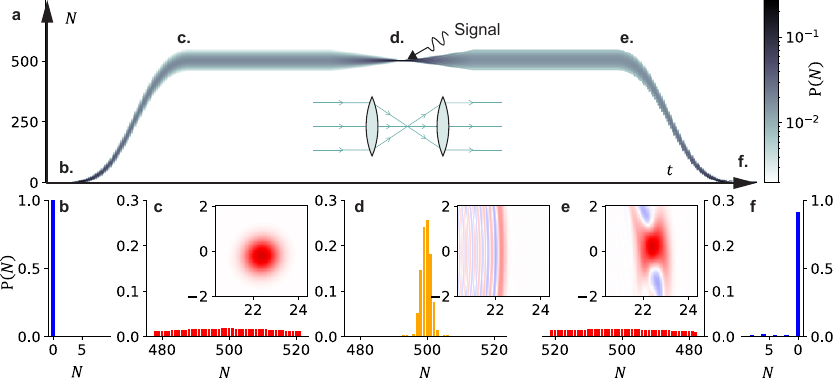}
    \caption{
    \textbf{Fock-space confocal system for state preparation and detection. }
	\textbf{a,} Changes of photon-number distribution during the Fock-space confocal circuit. The circuit maps a coherent state back to an approximate coherent state using two Fock-space convex lenses with the same focal point, analogous to the confocal convex lens system in optics. \textbf{b-f,} Close-up of photon-number distribution at various moments. \textbf{b,} At the starting point, the system is in the vacuum state, with all population being in the Fock state $\ket{0}$. \textbf{c,} A displacement transformation maps the vacuum state to a coherent state with 500 average photons and a broad distribution. The inset shows the Wigner distribution of the coherent state. \textbf{d,} After passing through a Fock-space convex lens, the coherent state is focused into an \textit{N}-focused state, with photon-number distribution tightly concentrated around several Fock states near $N=500$. The inset shows the Wigner distribution of the \textit{N}-focused state, which has fine fringes for detecting small signals. \textbf{e,} After passing through another Fock-space convex lens, the \textit{N}-focused state diverges into a coherent-like state, with a broad photon-number distribution. The inset shows its Wigner distribution, which is generally similar to that of a coherent state except for slight distortions. \textbf{f,} A final backward displacement transformation maps the coherent-like state to a near vacuum state. Theoretically, over 90\% of the population returns to the vacuum state, facilitating efficient state measurement.
    }
    \label{fig1}
\end{figure*}

%Third paragraph: In this work
Here, we propose and experimentally demonstrate ``quantum  confocal microscopy", a paradigm that simultaneously resolves the challenges in efficient generation and detection of certain non-classical states in high-dimensional Hilbert spaces. Drawing a direct analogy between classical wave optics and quantum state evolution in a bosonic mode, we construct a 4f confocal system using lenses in Fock space. This architecture deterministically focuses a readily generated large coherent state into a non-classical Fock-space-focused (\textit{N}-focused) state characterized by sharply compressed photon-number uncertainty~\cite{Yifang2026FockOptics,Ming2026Fock10000}. The confocal microscopy is inherently scalable with a constant operational depth $\mathcal{O}(1)$. We experimentally prepare \textit{N}-focused states with a mean photon number up to 500 and a central Fock-state population surpassing 20\%. Crucially, by leveraging the confocal architecture, we achieve approximate time-reversal via a secondary Fock-space convex lens, remapping the \textit{N}-focused state near a coherent state for simple and efficient information extraction. This deterministic confocal metrology circuit achieves an information extraction efficiency exceeding 90\%, showing a quantum-enhanced metrological gain of 19.06~dB beyond SQL. Our work establishes a scalable framework for preparing large-scale non-classical states, marking a critical step toward quantum-enhanced sensing technologies.

\section{Results}
\subsection{Confocal system in Fock space}

% principle of the scheme
% why \textit{N}-focused state
\noindent Figure~\ref{fig1}(a) illustrates the concept of quantum confocal microscopy in Hilbert space. It shows the evolution of a bosonic mode quantum state throughout the quantum circuit, with the state distribution along the photon-number basis. The inset depicts the corresponding classical 4f confocal system widely used in optical microscopy, which consists of two lenses with coincident focal points. The first lens transforms a broad collimated beam into a tight, diffraction-limited focal spot, while the second lens re-collimates the beam, enabling high resolution and efficient signal collection. Building on an equivalence between optical wave propagation in real space and quantum state evolution in Fock space (\red{Supplementary Section I A})~\cite{Yifang2026FockOptics,Ming2026Fock10000}, coherent states as a Gaussian beam in Fock space can be focused into \textit{N}-focused states and subsequently remapped back to a coherent state near the vacuum (\red{Supplementary Section I B}).

% why confocal system

\begin{figure*}
    \includegraphics{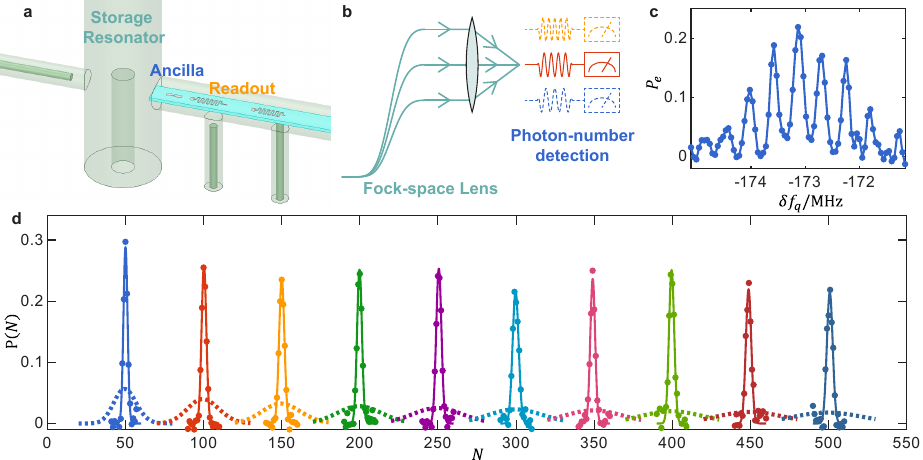}
    \caption{
    \textbf{Scaling of \textit{N}-focused state.}
    \textbf{a,} Schematic of the superconducting circuit QED setup: a 3D cavity coupled to a transmon qubit and a readout resonator. 
    \textbf{b,} The circuit of preparing and calibrating the \textit{N}-focused states. The \textit{N}-focused state is prepared by compressing a coherent state through a convex Fock-space lens, then calibrated with a photon number selection pulse and measurement of the transmon qubit.
    \textbf{c,} Ancilla-qubit-assisted detection results for an \textit{N}-focused state with an average photon number of $\bar{N}=300$. 
    \textbf{d,} Measured $P(N)$ (dots) and Gaussian fits (solid lines) for the \textit{N}-focused state at various $\bar{N}$ from 50 to 500, demonstrating deep sub-Poissonian statistics compared to coherent states (dashed lines). The peak probability and the width of the distribution remain nearly the same at various target photon numbers, confirming the scalability of the state preparation process. 
    } 
    \label{fig2}
\end{figure*}

% circuit
The principle of the quantum confocal microscopy is numerically simulated and illustrated in Figs.~\ref{fig1}(b-e), with a constant operation depth independent of $N$. Initially, the bosonic mode resides in the vacuum state, with its photon-number distribution shown in Fig.~\ref{fig1}(b). A large displacement operation first converts the vacuum state into a coherent state with an average photon number $\bar{N}=500$, with photon-number distributions and Wigner functions depicted in Fig.~\ref{fig1}(c). Subsequently, a Fock-space convex lens, composed of sequential quadratic phase accumulation followed by weak-driving evolution, transforms the coherent state into an \textit{N}-focused state, as shown in Fig.~\ref{fig1}(d). A second lens with identical parameters and a common focal point then reconverts the \textit{N}-focused state into a quasi-coherent state, as shown in Fig.~\ref{fig1}(e). These numerical results validate bidirectional state conversion between coherent and \textit{N}-focused states via the confocal architecture, confirming our optical intuition. The photon-number distribution of the quasi-coherent state closely resembles the coherent state, while minor distortions in the Wigner function arise from the inherent discrete nature of Fock-space (\red{see Supplementary Section I C}). Finally, a reverse large displacement operation repositions the quasi-coherent state near the vacuum origin, with the resulting photon-number distribution shown in Fig.~\ref{fig1}(f).

The \textit{N}-focused state exhibits the fundamental quantum enhancement for metrological gain beyond SQL. As shown in Fig.~\ref{fig1}(d), experimental feasible parameters allow compression of the photon-number distribution to a narrow full width at half maximum (FWHM) of around 4 photons for $\bar{N}=500$. In particular, the population on the Fock state $\ket{N=500}$ with a deterministic photon number exceeds 20\%. 
The Wigner distribution of the \textit{N}-focused state (inset of Fig.~\ref{fig1}(d)) reveals dense, ring-like fringes analogous to those of Fock states, confirming the high-precision metrological capability~\cite{Zurek2001SubPlanck}, in contrast to the coherent state in Fig.~\ref{fig1}(c). Moreover, the negativity of the Wigner distribution demonstrates that the $N$-focused state is non-Gaussian, distinguishing it from conventional squeezed states. 
%Conceptually, a sharper focus in the Fock basis (i.e., narrower photon-number distribution) corresponds to finer phase space fringes, perfectly analogous to a classical confocal microscope, where a tighter focal spot enhances spatial resolution. This delicate structure is exceptionally sensitive to external perturbations like phase space displacement, making the \textit{N}-focused state an ideal quantum sensing probe.

%We employ the \textit{N}-focused states, composed of several adjacent Fock-state components, as probes for precision metrology. While large Fock states are well-known for their precision measurement capabilities, their deterministic preparation remains experimentally challenging. In contrast, \textit{N}-focused states can be efficiently generated using recently developed Fock-space optical methods~\cite{Yifang2026FockOptics,Ming2026Fock10000}. Crucially, these \textit{N}-focused states exhibit intrinsic metrological capability: their QFI scales linearly with average photon number, though it remains lower than that of ideal Fock states (\red{Supplementary Section IV A}). This property facilitates direct utilization of \textit{N}-focused states as measurement probes, circumventing complex state-preparation protocols and unlocking scalable quantum-enhanced metrological gains.

Another key advantage of this confocal design is that it resolves the information readout bottleneck. Rather than directly measuring the macroscopic post-sensing state, this approach employs a second Fock-space lens followed by a reverse displacement to implement an approximate time-reversal. 
This maps the complex state imprinted with the test perturbation back near the vacuum state $\ket{0}$, converting the tiny shift at the ``focus" into a simple observable change near vacuum. 
Simulations show that over 90\% of the population returns to the vacuum state in the absence of test perturbation, correlating the magnitude of mid-circuit perturbations with the change in the final vacuum-state fidelity and enabling precise signal detection. Thus, information is extracted via efficient near-vacuum measurements, circumventing the need for macroscopic quantum state tomography. 
Crucially, unlike conventional time-reversal protocols that require precise Hamiltonian inversion ($H \rightarrow -H$)~\cite{Davis2016Approaching,Colombo2022Time-reversal,Zeyang2023quantumscramblingmetrology}, our protocol, analogous to classical optics, requires no such inversion and is inherently robust.

\subsection{Characterization of the Fock lens and confocal system}

% (1) describe the physical system
\noindent Our circuit quantum electrodynamics (QED) system~\cite{Wallraff2004Strongcoupling,Koch2007Transmon,Blais2021Circuit} consists of three key components: a high-quality memory resonator stores quantum information, an ancilla transmon qubit, and a readout resonator, as shown in Fig.~\ref{fig2}(a). 
The storage resonator features a long lifetime of 2.8~ms and acquires Kerr nonlinearity via coupling with the ancilla qubit. 
The dispersive coupling between the ancilla qubit and the storage resonator causes frequency shifts on the qubit when the storage resonator is in specific Fock states. Therefore, applying frequency-selective drives on the qubit can distinguish different Fock states in the resonator~\cite{Schuster2007NumSpl}. 
The readout resonator, with a short lifetime ($ < 1\,\mu$s), is also dispersively coupled to the ancilla qubit. We measure the state of the ancilla qubit by performing homodyne detection on the readout resonator. Detailed system parameters and wiring are given in \red{Supplementary Section II}. 

% (2) show the photon number dynamics
% Figure 2b provides a temporal snapshot of the core confocal dynamics, visualizing the evolution of the photon-number distribution, $P(n)$. We initialize the system in a large coherent state (e.g., at $\langle N \rangle \approx N$) with its characteristic broad, Poissonian distribution (left panel), analogous to a collimated beam. As the Fock lens Hamiltonian is applied, the distribution undergoes a focusing evolution. The $P(n)$ rapidly contracts, collapsing at the focal time $t_f$ into a single, sharply peaked, subPoissonian distribution (center panel), which is our target probe state. Crucially, we demonstrate that by letting the evolution continue past this focal point ($t > t_f$), the state naturally diverges the distribution broadens again (right panel), effectively undoing the compression. This deterministic focusing and diverging evolution is the temporal realization of the confocal optical path and forms the physical basis for our approximate time reversal readout protocol.

%(2) qubit-assisted photon number detection
%We start by calibrating the \textit{N}-focused state with qubit-assisted photon number detection. 
The photon-number distribution of the \textit{N}-focused state is calibrated by qubit-assisted photon-number detection.
We first prepare an \textit{N}-focused state in the storage cavity using the circuit illustrated in Figs.~\ref{fig1}(b-d). Then, frequency-selective $\pi$-pulses are applied to the ancilla qubit at varying frequencies, and successful qubit flips are identified by reading out the ancilla qubit through the readout resonator, as shown in Fig.~\ref{fig2}(b). Figure~\ref{fig2}(c) presents the detection results for an \textit{N}-focused state with an average photon number of $\bar{N}=300$. The qubit excitation is detectable only when the cavity has population in a specific Fock state, and the qubit drive frequency aligns with the transition frequency corresponding to that Fock state. By systematically preparing \textit{N}-focused states with different average photon numbers and analyzing the qubit excitation peak positions one by one, we identify the transition frequencies of specific Fock states within the cavity (\red{Supplementary Section II D}). Furthermore, applying a set of distinct qubit drive frequencies to a given \textit{N}-focused state allows quantification of the population distribution across its Fock state components.

% (4) show the scalability of scheme
%A critical figure of merit for any practical quantum protocol is its scalability. We investigate the prepared \textit{N}-focused states as the target photon number $\bar{N}$ increases. The dots in Fig.~\ref{fig2}(d) demonstrate the experimentally measured photon-number distributions of the \textit{N}-focused states with varying average photon numbers. 
We demonstrate the scalability of the Fock-space focusing circuit by measuring the photon-number distributions of the \textit{N}-focused states at various average photon numbers $\bar{N}$, as shown in Fig.~\ref{fig2}(d).
The solid lines represent Gaussian fits to the experimental data, while the dashed lines correspond to coherent states with matching $\bar{N}$. As $\bar{N}$ increases, the prepared \textit{N}-focused state distributions remain essentially stable, whereas the coherent state distributions broaden proportionally to $\sqrt{\bar{N}}$. We fit the distribution of the \textit{N}-focused state with Gaussian function $P(N) = A\cdot \exp{\left(-(N-\bar{N})^2/(2\sigma^2)\right)}$, as shown by the dashed lines in Fig.~\ref{fig2}(d). The fitted standard deviation $\sigma$ is smaller than 2 photons for all $\bar{N}$, consistent with the theoretical prediction (\red{Supplementary Section IV A}). At $\bar{N} = 500$, the central occupation of the \textit{N}-focused state remains above 20\%, with a fitted $\sigma=1.9\pm0.3$ photons, corresponding to a 21.5$\pm$1.1~dB compression compared to coherent states. However, the measurement fidelity of the ancilla qubit degrades with increasing photon numbers, resulting in elevated noise levels in the experimental data (\red{Supplementary Section II D}). The limited fidelity of direct measurement not only restricts potential post-measurement selection operations, but also limits the precision of directly applying \textit{N}-focused states for quantum metrology.

% (3) highlight the photon-number distribution for Fock 500
% The "focal points" of our system are deeply \textit{N}-focused states with giant photon numbers. In Figure 2c, we present the experimentally measured photon-number distributions, $P(n)$, for three distinct probe states, 'focused' at target numbers $N \approx 100$, $300$, and $500$. Each distribution is exceptionally narrow, demonstrating profound non-classicality. For our largest state, $N \approx 500$, the population is tightly confined, with the peak probability $P(n=500)$ exceeding 20\%. To highlight the degree of compression, we superimpose the $P(n)$ of a classical coherent state (dashed lines) with the same mean photon number. The stark contrast confirms the powerful sub-Poissonian nature of our deterministically prepared states, with a photon number squeezing ratio exceeding $xx$dB.

\begin{figure*}
    \includegraphics{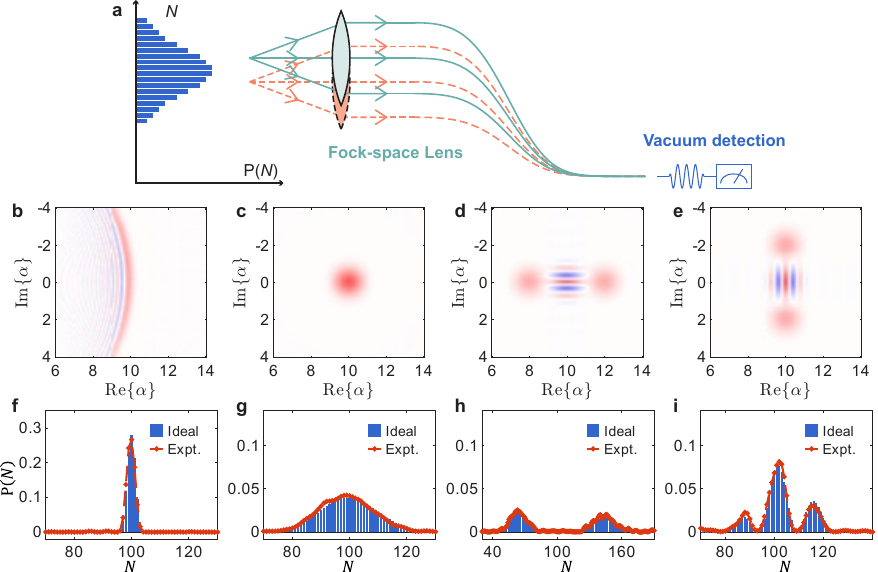}
    \caption{\textbf{Fock-space scanning tomography (\textit{N}-tomo) of various states.}
    \textbf{a,} The circuit of calibrating an unknown target state with \textit{N}-tomo. The target state passes through a Fock-space convex lens and then a displacement operation, followed by measuring the vacuum probability of the final state. The \textit{N}-tomo circuit maps the \textit{N}-focused state component in the target state back to the vacuum state, similar to the optical conjugation of the \textit{N}-focused state preparation circuit. By adjusting the circuit parameters, the \textit{N}-tomo circuit can detect the proportions of different \textit{N}-focused components in the target state, thereby reconstructing the photon-number distribution. 
    \textbf{b-e,} The theoretical Wigner distribution of the \textit{N}-focused state, the coherent state, the parallel-displaced cat (PDC) state, and the orthogonal-displaced cat (ODC) state. \textbf{f-i,} The ideal and reconstructed photon-number distribution of each state. The distribution of the \textit{N}-focused state is used for reconstructing other unknown states. The PDC state has two components with different photon numbers, resulting in a two-peak photon-number distribution. The ODC state has two components with the same photon numbers, resulting in a three-peak photon-number distribution. The \textit{N}-tomo results clearly capture the difference between PDC and ODC states.
     }
    \label{fig3}
\end{figure*}

% (4) Fock-space tomography principle
As a reversal of the preparation of \textit{N}-focused state, the Fock-space lens can be applied to detect the projection of state on a \textit{N}-focused state.  Figure~\ref{fig3}(a) shows detection channels with varying mean photon numbers, which could serve as basis vectors for quantum state tomography, enabling precise photon-number distribution measurement of an input state, termed Fock-space scanning tomography (\textit{N}-tomo). In this scheme, an arbitrary initial state undergoes a convex lens and displacement operation. The \textit{N}-focused state component within the initial state is mapped near the vacuum state, allowing quantification of the \textit{N}-focused component fraction in the initial state by measuring the vacuum-state proportion \red{(Supplementary Section III A)}.

% (5) Fock-space tomography result
We experimentally demonstrate the reconstruction of photon-number distributions from \textit{N}-tomo results for various initial states, along with their corresponding theoretical distributions and Wigner distributions, as shown in Figs.~\ref{fig3}(b-i). 
%For different initial states, the \textit{N}-tomo results differ from the theoretical photon-number distributions by only a constant scaling factor. 
We first reconstruct the photon-number distribution of the \textit{N}-focused state using its \textit{N}-tomo results, as shown in Figs.~\ref{fig3}(b-c). Based on the reconstructed distribution of the \textit{N}-focused state, we subsequently reconstruct the photon-number distributions of other unknown pure states from their \textit{N}-tomo results, with detailed methods provided in the \red{Supplementary Section III D}.
%Theoretical analysis reveals that \textit{N}-tomo is equivalent to the Q-function along a specific axis, enabling the inversion of photon-number distributions from \textit{N}-tomo measurements, \red{see supplementary material}. 

Specifically, we apply \textit{N}-tomo to the coherent state $ \mathcal{D} [\sqrt{n_t}] \ket{0} $, parallel-displaced cat (PDC) state $ \mathcal{D} [\sqrt{n_t}] ( \ket{\alpha} + \ket{-\alpha} ) / \sqrt{2} $ and orthogonal-displaced cat (ODC) state $ \mathcal{D} [\sqrt{n_t}] ( \ket{i\alpha} + \ket{-i\alpha} ) / \sqrt{2} $. The state preparation protocol is detailed in \red{Supplementary Section III B}. We choose $ n_t = 100 $ and $\alpha = 2$ as an example.
The coherent state has a wide Gaussian-like photon-number distribution centered at $ n_t = 100 $, as shown in Figs.~\ref{fig3}(d-e). 
The PDC state has two coherent state components with significantly different average photon numbers (64 and 144 photons), as shown in Figs.~\ref{fig3}(f-g). The ODC state consists of two coherent states with identical average photon numbers (104 photons) but a relative phase difference, resulting in a characteristic triple-peak structure in the photon-number distribution, as shown in Figs.~\ref{fig3}(h-i). Experimentally reconstructed results for all types of states agree well with theoretical predictions, confirming that \textit{N}-tomo not only possesses a broad detection range but also exhibits high resolution for fine structures in photon-number distributions arising from coherent superpositions.

\subsection{Quantum sensing of displacement}

\begin{figure*}
    \includegraphics{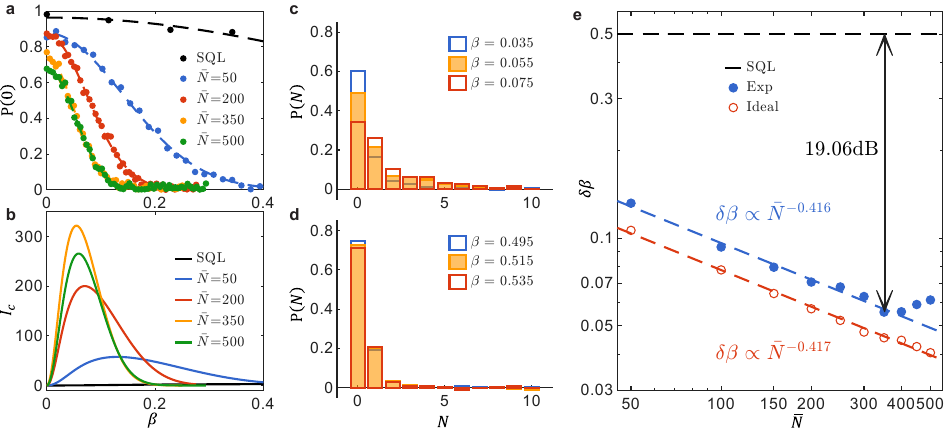}
    \caption{\textbf{Quantum sensing of displacement.} 
    \textbf{a,} Measured vacuum population $P(0)$ vs. displacement $\beta$. \textit{N}-focused states ($\bar{N}$ from 50 to 500) show a steeper slope (higher sensitivity) than the coherent state (SQL benchmark). 
    \textbf{b,} Fisher Information $I_c$ (derived from \textbf{a}) shows a massive quantum enhancement for \textit{N}-focused states over the coherent state at optimal bias points $\beta_0$. 
    \textbf{c,} Focused $\bar{N}=350$ state detection result. The same signal $\delta\beta$ (bottom) induces a dramatic, measurable change in the photon-number distribution of the final state, illustrating the mechanism of quantum gain. 
    \textbf{d,} Coherent state (SQL) detection result. A small signal $\delta\beta$ (bottom) causes a negligible change to the final state. \textbf{e,} Displacement sensitivity $\delta\beta$ versus $\bar{N}$. The \textit{N}-focused states (blue dots) demonstrate a clear quantum advantage, surpassing the SQL ({black dashed line}). A maximum quantum gain of $19.06\pm0.13$ dB is achieved at $\bar{N} = 350$. Linear fitting (blue dashed line) reveals that the sensitivity scales as $\bar{N}^{-0.416}$, approaching the Heisenberg-limit scaling ($N^{-0.5}$). This scaling is well-described by lossless simulation (yellow circles and dashed line, $\bar{N}^{-0.417}$).
    %The experimental scaling (dashed line, $\bar{N}^{-0.416}$) demonstrates a near-Heisenberg-limit scaling ($N^{-0.5}$)
    %approaches the ideal Heisenberg limit ($N^{-0.5}$) and is well-described by our theoretical model \red{(solid line)}. 
    }
    \label{fig4}
\end{figure*}

% sensing protocol and data processing
\noindent We now integrate the preparation and detection protocols of \textit{N}-focused states to achieve quantum-enhanced metrology. The protocol employs a confocal system where a small, unknown displacement $D(\beta)$ is applied at the focal point.
%followed by detection of the near-vacuum final state across the entire detection pathway. 
The displacement magnitude information is extracted by measuring the final vacuum-state probability $P(0)$.
To calibrate the detector response, we first apply a known displacement $\beta_0$ and scan its intensity. Figure~\ref{fig4}(a) demonstrates the experimental measurement results of $P(0)$, with distinct color points representing results for different average photon numbers ($\bar{N} = 50, 200, 350, 500$). Black points represent coherent-state probe measurement results, corresponding to the SQL in our system. Notably, the confocal system demonstrates a significantly steeper slope than that of a coherent state. 

To extract the information in the measurement result, we first fit the experimental data with a {Gaussian function $P(0|\beta) = A\cdot \exp{\left(-\beta^2/(2\sigma^2)\right)}+C$} (dashed lines in Fig.~\ref{fig4}(a)). The fitting curves agree well with the experimental data. We then calculate the classical Fisher information (CFI) using the formula $I_c(\beta) = (\partial P(0|\beta)/\partial\beta)^2 /(P(0|\beta)(1-P(0|\beta)))$, and the results are shown in Fig.~\ref{fig4}(b). The results show that $I_c(\beta)$ for our confocal metrology circuit peaks at specific optimal bias points $\beta_0$ for various average photon numbers $\bar{N}$. At $\bar{N}=350$, the maximum CFI reaches $3.22\pm0.10\times10^2$ at $\beta_0=0.055$, surpassing the ideal SQL ($I_c = 4$) by a factor of 80.5.
%, corresponding to a metrology gain of 19.06dB. 

% final state P_n
To intuitively visualize the origin of sensing enhancement, we characterize the population distribution of the lowest few Fock components in the final output states, as shown in Fig.~\ref{fig4}(c). 
We select the operating point with the highest precision at $\bar{N}=350$ and bias the system at $\beta_0 = 0.055$, where the CFI is maximized. 
The populations of Fock components with photon numbers $N \leq 10$ are presented as solid bars in Fig.~\ref{fig4}(c). When a test drive differs from the bias point by $\Delta\beta = \pm0.02$, the changed Fock populations are shown as hollow bars in Fig.~\ref{fig4}(c). The vacuum-state ($\ket{0}$) population exhibits a significant variation when the test drive is applied. 
For comparison, Fig.~\ref{fig4}(d) shows the final state population changes of a coherent-state probe at its optimal bias $\beta_0=0.515$. Under the same test drive strength, the variation of the final state in the confocal metrology circuit is remarkably larger than that of the coherent-state circuit. 
Due to system imperfections such as aberrations and dissipation, not all population returns to the vacuum state after one confocal circuit, resulting in a gradually decreasing distribution across the Fock states $\ket{1}$ to $\ket{10}$ and a residual population in states $\ket{N>10}$. We observe that the populations in Fock states $\ket{1\leq N\leq10}$ also vary with the test drive strength, suggesting that quantum information might be partially recovered via error correction or mitigation techniques. For instance, one can construct low-Fock-number operations with numerically optimized pulse shaping to transfer more population on the Fock states $\ket{1\leq N\leq10}$ back to $\ket{0}$, thereby enhancing the metrology performance.

% scaling of metrology gain
% We now present the scalability of our confocal circuit and the resulting metrological gain. 
According to the Cram\'{e}r-Rao theorem, the minimum achievable uncertainty $\delta\beta$ for the test small displacement $\beta$ can be directly obtained from the CFI of the measurement result, i.e., $\delta\beta = 1/\sqrt{I_c}$.
Figure~\ref{fig4}(e) plots the measured displacement uncertainty, $\delta\beta$, as a function of the mean photon number $\bar{N}$ on a log-log scale. The sensitivity improves rapidly as $\bar{N}$ increases.
%The experimentally measured $|\delta\beta|$ lower bounds of the confocal circuit at different average photon numbers are shown by the data points in Fig.~\ref{fig4}(e), 
% a clear and rapid improvement in precision (a decreasing $|\delta\beta|$) as $\bar{N}$ increases.
For comparison, an ideal coherent state can only yield $|\delta\beta| \geq 0.5$ regardless of the number of photons used, representing the SQL as $\delta\beta_\mathrm{SQL}=0.5$, as shown by the black dashed line in Fig.~\ref{fig4}(e).
The ratio $ 10\log_{10}\left(| \delta\beta_{\text{SQL}} / \delta\beta | )^2 \right)$ defines the quantum metrology gain in decibels. The minimum $|\delta\beta|$ measured experimentally occurs at $\bar{N}=350$, where $|\delta\beta| \geq 0.0557$, corresponding to a $19.06\pm0.13~\mathrm{dB}$ gain over coherent states. Further increasing the photon number leads to an increase in $|\delta\beta|$, as the benefits of increased $\bar{N}$ are gradually overcome by the increased photon-loss rate within the high-photon-number range. 

By fitting the experimental data for $\bar{N} < 350$ and the simulation data, we find that the measurement precision varies with the average photon number as $\delta\beta \propto \bar{N}^{-0.416}$. 
%(experiment) and $\delta\beta \propto \bar{N}^{-0.417}$(lossless simulation), respectively.
This result is a clear demonstration of near-Heisenberg-limited (HL) scaling, which for an ideal Fock state probe is $\delta\beta_{\text{HL}} \propto N^{-0.5}$. 
We also numerically simulate the maximum precision of the confocal circuit under lossless conditions, as shown by the circles in Fig.~\ref{fig4}(e). The simulated measurement precision scales with $\bar{N}$ as $\delta\beta \propto \bar{N}^{-0.417}$, which agrees well with the experimental result. The lossless simulation indicates that the measurement precision can be further improved by extending the cavity lifetime. Our results confirm that the quantum confocal microscop in Fock space is highly effective and scalable.

%In Figure 4a, we first quantify the performance of the 'Fock lens' itself by plotting the photon number squeezing ratio, defined as the variance of a classical coherent state ($\sigma^2_{\text{coh}} = \langle N \rangle$) relative to the measured variance of our squeezed state ($\sigma^2_{\text{sqz}}$). This metric, $\sigma^2_{\text{coh}} / \sigma^2_{\text{sqz}}$, directly measures the 'focusing' efficiency. The experimental data demonstrate a significant and robust squeezing factor across the entire range, from small $N$ up to $N=500$. The results show excellent agreement with our theoretical model (solid line), confirming that our 'focusing' protocol is highly effective and scalable.

%This remarkable performance is contingent on the 'focusing' condition. The maximum gain is achieved only when the lens achieves its "diffraction limit". The optimal focusing condition, $\phi_0 \sqrt{N} \sim 1$, ensures the narrowest focus point and thus the highest sensitivity, further reinforcing the deep and powerful connection between the principles of classical confocal optics and the ultimate performance of our quantum metrological system.

\section{Discussion}

% Conclusion
\noindent In this work, we have introduced the concept of quantum confocal microscopy in Fock space and demonstrated it experimentally as a unified framework for the deterministic preparation, manipulation, and detection of non-classical bosonic states at macroscopic photon numbers. In a superconducting circuit QED platform, merely requiring Kerr nonlinearity and weak driving, we have successfully realized a 4f confocal microscopy and prepared \textit{N}-focused states with up to 500 photons, demonstrating the scalability of the approach. Utilizing the latter half of the confocal system, we implement Fock-space scanning tomography to detect the photon-number distribution of unknown states, achieving excellent agreement with theory. Finally, we employ the complete confocal circuit for displacement sensing, achieving a quantum-enhanced precision exceeding the SQL by 19~dB. 
Our experiment provides a scalable method for leveraging high-photon-number states in quantum metrology. 
Requiring only the elementary nonlinearity of a bosonic resonator and vacuum-state detection, our approach is highly adaptable to a wide range of other bosonic architectures, such as acoustic~\cite{Chu2018acousticFock}, optomechanical~\cite{Barzanjeh2022Optomechanics}, trapped ions~\cite{Wolf2019MotionalFock}, and cavity QED systems~\cite{Blais2004CavityQED,Wallraff2004Strongcoupling,Yang2023Non-Reciprocal}, for quantum-enhanced sensing of diverse physical parameters.

%The results also demonstrate the effectiveness of Fock-space optical methods for both controlling high-photon-number states and realizing quantum-enhanced precision measurement. 
%\textcolor{cyan}{The confocal system metrology framework is readily applicable to other high-dimensional quantum systems, such as quantum mechanical oscillators~\cite{Chu2018acousticFock}, motional modes in ion traps~\cite{Wolf2019MotionalFock}, and photonic systems~\cite{Pirandola2018photonicquantumsensing}.}

% error analysis
The primary source of dissipation in this experiment is single-photon loss in the high-photon-number range. Due to the bosonic nature of photons, the lifetime of a Fock state $\ket{N}$ scales inversely with photon number, $ T_1 \propto 1/N $. 
However, the phase accumulation time of the confocal lens system is determined by the paraxial condition (Supplementary Section I) and scales as $T_L \propto 1/\sqrt{\bar{N}}$~\cite{Ming2026Fock10000}. Consequently, the total execution time of the confocal circuit decreases with $\bar{N}$ for a fixed Kerr nonlinearity, partially mitigating the impact of single-photon dissipation. We anticipate that even higher metrological gain is possible by further optimizing the lifetime of the cavity~\cite{Milul2023Cavity34ms}.
In experimental systems, the Kerr nonlinearity strength often depends on $\bar{N}$. Designing specialized nonlinear devices to maintain or even enhance the Kerr nonlinearity in the high-photon-number range~\cite{Hua2025NEMS} offers a promising path toward improving circuit efficiency and metrological performance.

\smallskip{}

%\noindent \textbf{\large{}Online content}{\large\par}
%\noindent Any methods, additional references, Nature Research reporting summaries, source data, extended data, supplementary information, acknowledgements, peer review information; details of author contributions and competing interests; and statements of data and code availability are available on line.

\bibliographystyle{Zou}
\bibliography{Library_for_Paper_hzy-new_v2}

%\clearpage{}
%\setcounter{figure}{0} 
%\noindent \textbf{\large{}Methods}{\large\par}

\vbox{}

\smallskip{}

%\noindent \textbf{\large{}Data availability}{\large\par}

%\noindent All data generated or analyzed during this study are available within the paper and its Supplementary Information. Further source data will be made available on reasonable request.

%\smallskip{}

%\noindent \textbf{\large{}Code availability}{\large\par}
%\noindent The code used to solve the equations presented in the Supplementary Information will be made available on reasonable request.

\smallskip{}

\noindent \textbf{\large{}Acknowledgment}{\large\par}

\begin{acknowledgments}
\noindent This work was funded by the National Natural Science Foundation of China (Grants No. 12550006, 92265210, 92365301, 92565301, 92165209, 12547179, 12574539, 12474498, 12404567, and 12504580) and the Quantum Science and Technology-National Science and Technology Major Project (2021ZD0300200 and 2024ZD0301500). This work is also supported by the Fundamental Research Funds for the Central Universities, the USTC Research Funds of the Double First-Class Initiative, the supercomputing system in the Supercomputing Center of USTC, and the USTC Center for Micro and Nanoscale Research and Fabrication.
%and the Guangdong Provincial Quantum Science Strategic Initiative (Grant No.~GDZX2403008, GDZX2503001). X.Z. acknowledges support from the Chinese Ministry of Science and Technology (MOST) through grant 2023ZD0300600, from the Hong Kong Research Grant Council through grant 17310725, and from the State Key Laboratory of Quantum Information Technologies and Materials. 
\end{acknowledgments}

\smallskip{}

\noindent \textbf{\large{}Author contributions}{\large\par}

\noindent
Z.H., M.L., and C.-L.Z. conceived the experiment and provided theoretical support.
Z.H., C.M., and Y.Z. performed the experiment, analyzed the data, and carried out the numerical simulations under the supervision of L.S.
Y.X., Z.C., and W.C. provided theoretical support.
J.C. helped to calibrate the system.
Y.X., L.X., H.H., and W.W. contributed to experimental support.
Z.H. designed the 3D cavity. H.L. and H.W. fabricated the tantalum transmon qubits.
Z.H., C.M., Y.Z., M.L., C.-L.Z., and L.S. wrote the manuscript with input from all authors.
C.-L.Z. and L.S. supervised the project.

\smallskip{}

\noindent \textbf{\large{}Competing interests}{\large\par}

\noindent The authors declare no competing interests.

\smallskip{}

%\noindent \textbf{\large{}Additional information}{\large\par}

%\noindent \textbf{Supplementary information} The online version contains supplementary material.

\noindent \textbf{Correspondence and requests for materials} should be addressed to M.L., C.-L.Z., or L.S.

\clearpage{}

\end{document}